\documentclass[prd,twocolumn]{revtex4}%
\usepackage{amsfonts}
\usepackage{amsmath}
\usepackage{amssymb}
\usepackage{graphicx}%
\begin{document}
\title{Entanglement is protected by acceleration-induced transparency in thermal field}
\author{Yongjie Pan$^{1,2}$}
\author{Baocheng Zhang$^{1}$}
\email{zhangbaocheng@cug.edu.cn}
\author{Qingyu Cai$^{2,3,4}$}
\email{qycai@hainanu.edu.cn}
\affiliation{$^{1}$School of Mathematics and Physics, China University of Geosciences,
Wuhan 430074, China}
\affiliation{$^{2}$Center for Theoretical Physics, Hainan University, Haikou 570228, China}
\affiliation{$^{3}$School of Information and Communication Engineering, Hainan University,
Haikou 570228, China}
\affiliation{$^{4}$Peng Huanwu Center for Fundamental Theory, Hefei 230026, China}
\begin{abstract}
The acceleration-induced transparency (AIT) effect has been suggested recently
to amply the transition probability of the two-level detctor and offers a
potential avenue for the experimental detection of the Unruh effect. In this
paper, we explore the influence of the AIT effect on quantum entanglement
between two detectors accelerated in a thermal field background, since the
thermal backgound field cannot be avoided completely in any experiments.
Interestingly, we find that although the backgound thermal field generally
degrade the entanglement between the detectors, the AIT effect can effectively
protect it.

\end{abstract}
\maketitle

\section{Introduction}

The Unruh effect predicts that an uniformly accelerated observer perceives the
Minkowski vacuum as a thermal bath with temperature $T=a/2\pi$, where $a$
denotes the proper acceleration \cite{unruh1976notes}. This remarkable effect
reveals profound connections between quantum field theory and general
relativity, demonstrating the observer-dependent nature of quantum vacuum. Not
only does it validate nontrivial aspects of quantum field theory in
non-inertial frames, but also establishes a paradigmatic framework for
studying black hole thermodynamics (e.g., Hawking radiation), thereby
deepening our understanding of spacetime, gravity, and quantum theory.
However, experimental verification remains challenging since achieving the
required acceleration ($\sim10^{20}\,\mathrm{m/s^{2}}$) exceeds current
technological capabilities. This makes the search for feasible indirect
detection methods a central problem in contemporary research.

Recent studies suggest that acceleration-induced transparency (AIT) may
provide a pathway to detect the acceleration-dependent response of the
Unruh-DeWitt (UDW) detectors \cite{dewitt1979quantum,soda2022acceleration}.
For a detector undergoing non-uniform acceleration in a multi-particle field,
the transition from ground to excited state receives two distinct
contributions, the stimulated absorption (that is the energy exchange with
field quanta proportional to particle number) and the Unruh contribution (that
is the accelration-driven excitation inherent to non-inertial motion). The AIT
phenomenon occurs when the stimulated absorption is suppressed while the Unruh
term gets amplified by a factor of $(\langle n\rangle+1)$, where $\langle
n\rangle$ represents the mean particle number in the field. Crucially, for
sufficiently large $\langle n\rangle$, this amplification could render the
detector's acceleration response experimentally measurable in the near future.

The influence of uniformly accelerated motion on quantum entanglement has been
extensively studied in various contexts. Notable examples include the Unruh
effect~\cite{crispino2008unruh,wang2011multipartite,richter2015degradation}
and anti-Unruh effects~\cite{li2018would,pan2020influence,yan2024entanglement}%
, as well as scenarios in curved spacetimes such as near Schwarzschild black
holes~\cite{fuentes2005alice,martin2010unveiling}. Overall, testing the Unruh
effect through acceleration-induced changes in entanglement is still beyond
the reach of current experimental technology. However, since entanglement is
fragile, the degradation of entanglement due to acceleration or gravity is
indeed worth studying, especially in the presence of a thermal background
field. The AIT effect indicates that acceleration-induced physical effects can
be amplified in a thermal field. This motivates our investigation into how
entanglement between accelerated detectors is affected by AIT in a thermal
background---whether it leads to more severe degradation or provides a certain
degree of protection---is the main focus of our study. Our results reveal that
AIT not only facilitates experimental detection of UDW detector responses but
also protects entanglement between accelerated detectors.

For different field particle distributions, such as in squeezed states,
coherent states, and thermal states, the amplification effect of field
particles on UDW detector responses differs \cite{Pan2023Enhanced}. Moreover,
the thermal effects from background fields and the Unruh effect appear to
produce entirely different outcomes for quantum entanglement, with existing
discussions for uniform acceleration cases \cite{brenna2016anti,pan2021anti}.
We therefore investigate entanglement evolution under uniform acceleration
when AIT occurs. This is particularly relevant for thermal field states, where
detector entanglement is influenced by two thermal effects: the Unruh-induced
thermal effect, and the background thermal field effect. Both effects
typically degrade entanglement. However, the emergence of AIT suppresses
partial contributions from the thermal background, leading to qualitatively
different entanglement dynamics.

This paper is organized as follows. First, in Sec. II, We have systematically
reviewed the response of accelerated detectors interacting with multi-particle
fields. In Sec. III, we investigated the AIT phenomenon for fields prepared in
both number states and thermal states. Sec. IV presents our analysis of
entanglement dynamics, demonstrating how the AIT mechanism can protect
detector-detector entanglement. Furthermore, we examined the entanglement
degradation caused by thermal effects in background fields. Finally, we
present our conclusions in Sec. V. In this paper, we use units with
$c=\hbar=k_{B}=1$.

\section{The detector model}

The model of the UDW detector can be embodied by a pointlike two-level quantum
system. Its interaction with the background field is expressed by the
Hamiltonian \cite{soda2022acceleration,birrell1984quantum},
\begin{equation}
H_{int}=\lambda\chi(\tau)\hat{\mu}(\tau)\hat{\phi}(t(\tau),x(\tau)),\label{H}%
\end{equation}
where $\lambda$ is the coupling constant between the accelerated detector and
the scalar field, $\hat{\mu}(\tau)=e^{i\Omega\tau}\hat{\sigma}_{+}%
+e^{-i\Omega\tau}\hat{\sigma}_{-}$ represents the detector's monopole moment
with $\sigma_{\pm}$ being $SU(2)$ ladder operators. The detector possesses two
distinct energy levels denoted by the ground $|g\rangle$ and excited
$|e\rangle$ state, respectively, separated by an energy gap $\Omega$ in the
detector's rest frame. $\hat{\phi}(t(\tau),x(\tau))=\int dk\,\left(
e^{-i\omega_{\mathbf{k}}t+i\mathbf{k}\cdot\mathbf{x}}\hat{a}_{\mathbf{k}%
}+e^{i\omega_{\mathbf{k}}t-i\mathbf{k}\cdot\mathbf{x}}\hat{a}_{\mathbf{k}%
}^{\dagger}\right)  $ is the field operator in which $t(\tau)$, $x(\tau)$
represents the detector's trajectory. $\chi(\tau)$ is the switching function.
In this paper, we set $\chi(\tau)$ to be equal to $1$ without affecting our
conclusions. Restricting attention to the weak-coupling regime, the transition
amplitude for the process $|\psi_{i}\rangle\rightarrow|\psi_{f}\rangle$ is
given as,
\begin{equation}
\mathcal{A}_{i\rightarrow f}=\int\langle\psi_{f}|\hat{H}_{int}(\tau)|\psi
_{i}\rangle d\tau=\int dk\,\mathcal{A}_{i\rightarrow f}(\mathbf{k}),
\end{equation}%
\begin{equation}
\mathcal{A}_{i\rightarrow f}(\mathbf{k})=\lambda\int d\tau\langle\psi
_{f}|(e^{i\Omega\tau}\hat{\sigma}_{+}+h.c.)(e^{-i\omega_{\mathbf{k}%
}t+i\mathbf{k}\cdot\mathbf{x}}\hat{a}_{\mathbf{k}}+h.c.)|\psi_{i}\rangle,
\end{equation}
where $\mathbf{k}$ ($\omega_{\mathbf{k}}$) is the momentum (frequency) of a
field quantum at the proper time $\tau$. Note that in this paper, we only
consider the massless scalar field with the relation, $\omega_{\mathbf{k}}=k$.
Further, the amplitude can be expressed as
\begin{align}
\mathcal{A}_{i\rightarrow f}(\mathbf{k}) &  =\lambda\left[  I_{-}%
(\Omega,\mathbf{k})\langle\psi_{f}|\hat{\sigma}_{+}\hat{a}_{\mathbf{k}}%
|\psi_{i}\rangle+h.c.\right]  \nonumber\\
&  \quad+\lambda\left[  I_{+}(\Omega,\mathbf{k})\langle\psi_{f}|\hat{\sigma
}_{+}\hat{a}_{\mathbf{k}}^{\dagger}|\psi_{i}\rangle+h.c.\right]  .
\end{align}
Note that both rotating-wave terms (first line) and counter-rotating-wave
terms (second line) contribute in general, weighted by $I_{\pm}$
respectively,
\begin{equation}
I_{\pm}(\Omega,\mathbf{k})=\int d\tau\,e^{i\Omega\tau\pm ik^{\mu}x_{\mu}%
(\tau)},\label{Ipm}%
\end{equation}
where $k^{\mu}x_{\mu}(\tau)=\omega_{\mathbf{k}}t(\tau)-\mathbf{k}%
\cdot\mathbf{x}(\tau)$.

For a detector moving within a multi-particle field, the evolution of the
total system state is governed by the time-evolution operator. To first-order
perturbation theory, this operator expands as
\begin{equation}
U=I-i\int d\tau H_{I}(\tau)+O(\lambda^{2}).\label{eq:unitary_approx}%
\end{equation}
In the interaction picture, this evolution is described by
\cite{Pan2023Enhanced}
\begin{subequations}
\begin{align}
U|g\rangle|n_{k}\rangle= &  |g\rangle|n_{k}\rangle-i\sqrt{n_{k}}I_{-}%
|e\rangle|n_{k}-1\rangle\nonumber\\
&  -i\sqrt{n_{k}+1}I_{+}|e\rangle|n_{k}+1\rangle,\label{e1}\\
U|e\rangle|n_{k}\rangle= &  |e\rangle|n_{k}\rangle-i\sqrt{n_{k}+1}I_{-}^{\ast
}|g\rangle|n_{k}+1\rangle\nonumber\\
&  -i\sqrt{n_{k}}I_{+}^{\ast}|g\rangle|n_{k}-1\rangle,\label{e2}%
\end{align}
where $|n_{k}\rangle$ is the Fock state of the field that indicates the
presence of $n$ photons in the $k$ mode. In the Eqs. (\ref{e1}) and
(\ref{e2}), the second and third terms on the right side represent the
rotating-wave and counter-rotating-wave terms, respectively. For instance, the
second term in Eq. (\ref{e1}) signifies the absorption of energy by the
detector from the field, causing a transition from the ground state to the
excited state, commonly referred to as the stimulated absorption term
\cite{soda2022acceleration}. The third term in Eq. (\ref{e1}) accounts for the
contribution of the Unruh effect, which arises from the accelerated motion of
the detector within the photon field, known as the stimulated Unruh effect
term. Equation (\ref{e2}) can be interpreted in a similarly way.

From Eq. (\ref{e1}), the transition probability from the ground state
$|g\rangle$ to the excited state $|e\rangle$ is given by
\end{subequations}
\begin{equation}
P_{|g\rangle\rightarrow|e\rangle}=n_{k}|I_{-}(\Omega,\mathbf{k})|^{2}%
+(n_{k}+1)|I_{+}(\Omega,\mathbf{k})|^{2},
\end{equation}
where $I_{-}$ corresponds to the stimulated absorption term and $I_{+}$
represents the Unruh effect contribution. When the detector undergoes uniform
acceleration in the vacuum field $(n_{k}=0)$, its worldline follows
\begin{equation}
t(\tau)=a^{-1}\sinh(a\tau),\quad z(\tau)=a^{-1}\cosh(a\tau
),\label{acceleration word line}%
\end{equation}
and the detector's transition probability becomes:
\begin{equation}
P_{|g\rangle\rightarrow|e\rangle}=\sum_{k}|I_{(+,k)}|^{2}.
\end{equation}
Substituting the uniformly accelerated worldline of the detector into the
transition amplitude in Eq. (\ref{Ipm}), we obtain
\begin{equation}
P_{|g\rangle\rightarrow|e\rangle}=\frac{2\pi}{\Omega a(e^{\Omega/k_{B}T_{U}%
}-1)}%
\end{equation}
This corresponds to a Bose-Einstein distribution at a temperature
$T_{U}=a/2\pi$. This implies that the conventional Unruh effect, which arises
from a uniformly accelerating motion of the detector in vacuum, is equivalent
to the detector being immersed in a thermal bath at a temperature of $T_{U}$.

\section{Acceleration-induced transparency}

For fields containing particles, the field quanta amplify both the stimulated
absorption term by a factor of $\langle n\rangle$ and the Unruh term by
$(\langle n\rangle+1)$. Concurrently, Barbara \textit{et al.}
\cite{soda2022acceleration} proposed the AIT phenomenon. AIT occurs under
specific conditions where the stimulated absorption contribution to the
transition probability becomes vanishingly small ($I_{-}\rightarrow0$), while
the Unruh term remains non-zero ($I_{+}\neq0$). Mathematically, this is
characterized by the ratio $I_{-}/I_{+}\rightarrow0$. During AIT, only the
Unruh term contributes to the detector's transition probability, amplified by
$(\langle n\rangle+1)$, which facilitates experimental detection of detector
responses \cite{soda2022acceleration}.

From the transition amplitude (\ref{Ipm}), the distinction between stimulated
absorption and Unruh contributions primarily depends on the detector's
worldline. Defining $\pi(\tau)=k^{\mu}x_{\mu}(\tau)$ as the phase function
(where $\dot{\pi}(\tau)=constant$ represents inertial motion), we consider the
standard scenario for the Unruh effect with the uniformly accelerated
worldline in Eq. (\ref{acceleration word line}). Substituting this into the
Unruh term coefficient (\ref{Ipm}) yields
\begin{equation}
I_{+}(\Omega,\mathbf{k})=\frac{-i}{4a\pi\sqrt{\pi}}e^{-\frac{i}{a}}%
e^{i\frac{\Omega}{a}\ln\left(  \frac{-i}{a}\right)  }\Gamma\left(
-i\Omega/a\right)  ,
\end{equation}
where $\Gamma(z)$ denotes the complex gamma function
\cite{ahmadzadegan2018unruh,giacomini2022second}.

To realize the AIT effect, we choose a worldline with non-uniform
acceleration
\begin{equation}
\dot{\pi}(\tau):=k%
\begin{cases}
v_{0} & \tau<0\\
v_{0}+a_{1}\tau & \tau\in\lbrack0,T_{1})\\
v_{1}+a_{2}(\tau-T_{1}) & \tau\in\lbrack T_{1},T_{2})\\
v_{2} & \tau\geq T_{2}%
\end{cases}
. \label{two acceleration wl}%
\end{equation}
This worldline indicates: in the initial stage, the detector moves inertially
with speed $s_{0}=1-v_{0}$; then at time $T_{1}$ begins uniformly accelerated
motion with acceleration $a_{1}$ until the speed changes to $s_{1}=1-v_{1}$;
subsequently undergoes another uniformly accelerated motion with different
acceleration $a_{2}$; and finally moves inertially with speed $s_{2}=1-v_{2}$.

For massless scalar fields, this phase function $\pi(\tau)=k^{\mu}x_{\mu}$
simplifies to $\pi\left(  \tau\right)  =k^{0}\left(  x^{0}\left(  \tau\right)
-\sum_{i}x^{i}\left(  \tau\right)  \right)  $. Compared with the calculation
process of the Unruh effect, the most significant difference is that the
acceleration is non-uniform and includes phases of inertial motion, where the
acceleration process contains both deceleration and acceleration, ending with
inertial motion.

\begin{figure}[ptb]
\includegraphics[width=0.45\textwidth]{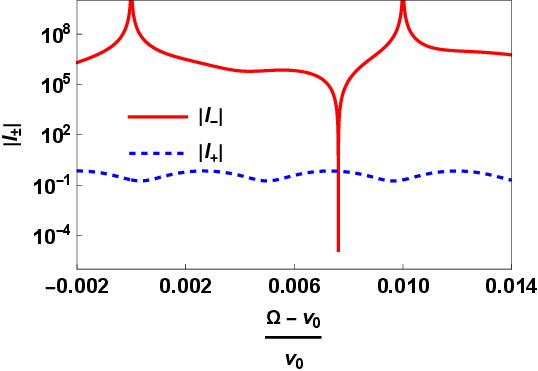} \caption{Acceleration-induced
transparency in thermal fields. The parameter is set to $\beta=0.1$,
$v_{0}=1.041$, $v_{1}=1.070$, $T_{1}=9.74350$, $T_{2}=1305.413$.}%
\label{Fig.1}%
\end{figure}

Since the thermal background cannot be neglected in actual experiments, we
consider AIT phenomenon for detectors moving in thermal fields which is
expressed in terms of the Fock state basis as
\begin{equation}
\rho_{k}=\sum_{n_{k}}(1-e^{-\beta\omega_{k}})e^{-n_{k}\beta\omega_{k}}%
|n_{k}\rangle\langle n_{k}|, \label{thermal state}%
\end{equation}
where $\beta=1/T$ is the inverse temperature of the background field. The
expectation value of the average particle number operator in the thermal state
is
\begin{equation}
\langle n_{k}\rangle=Tr(\rho_{k}n_{k})=\frac{1}{e^{\beta\omega_{k}}-1}.
\end{equation}

The transition probability for a detector moving in the thermal field is
calculated to be
\begin{equation}
P_{TS}=\frac{1}{e^{\beta\omega_{k}}-1}|I_{(-,k)}|^{2}+\left(  \frac
{1}{e^{\beta\omega_{k}}-1}+1\right)  |I_{(+,k)}|^{2},
\label{thermal state transition probability}%
\end{equation}
where the first term represents stimulated absorption while the second term
corresponds to the Unruh contribution. For the AIT effect to occur in the
thermal field, the stimulated absorption has to be suppressed, which can be
ensured by a small amplitude $|I_{-}|$ seen from $\frac{\text{Abs term}%
}{\text{Unruh term}}=e^{-\beta\omega_{k}}\frac{|I_{-}|^{2}}{|I_{+}|^{2}}$.

After calculation, we obtain the transition amplitude for an accelerating
detector in the thermal field, as shown in Fig. 1. It can be observed that the
contribution from the stimulated absorption process becomes significantly
smaller than that of the Unruh process at a specific energy gap where AIT
effect occurs.

\section{Entanglement}

Now we consider the evolution of quantum entangled states between two
detectors $A$ and $B$ along the worldlines which can lead to AIT. First,
assume the detectors' initial state
\begin{equation}
|\phi\rangle=(\alpha|g\rangle|e\rangle+\beta|e\rangle|g\rangle
),\label{entanglement state}%
\end{equation}
where $\alpha$ and $\beta$ are normalization coefficients satisfying
$\left\vert \alpha\right\vert ^{2}+\left\vert \beta\right\vert ^{2}=1$. For
fields prepared in the Fock states, the complete system's state evolution
follows $U_{A}^{\ast}U_{B}^{\ast}|\phi\rangle\langle\phi|U_{A}U_{B}$, where
the time evolution operators $U_{i},i\in\{A,B\}$ are given by Eq.
(\ref{eq:unitary_approx}).

The initial entangled state from Eq. (\ref{entanglement state}), moving along
worldline (\ref{two acceleration wl}) and interacting with the field
(particles), evolves into
\begin{equation}
\rho_{out}=%
\begin{pmatrix}
\rho_{gggg}^{11} & 0 & 0 & \rho_{eegg}^{41}\\
0 & \rho_{gege}^{22} & \rho_{egge}^{32} & 0\\
0 & \rho_{geeg}^{23} & \rho_{egeg}^{33} & 0\\
\rho_{ggee}^{14} & 0 & 0 & \rho_{eeee}^{44}%
\end{pmatrix}
\label{rho}%
\end{equation}
with the matrix basis $\{|gg\rangle,|ge\rangle,|eg\rangle,|ee\rangle\}$. And
each element is obtained as follows \begin{widetext}%
\begin{align*}
\rho_{gggg}^{11} &  =\alpha^{2}(\langle n_{B}\rangle|I_{+,B}|^{2}+(\langle
n_{B}\rangle+1)|I_{-,B}|^{2})+\beta^{2}\langle n_{A}\rangle|I_{+,A}|^{2}%
+\beta^{2}(\langle n_{A}\rangle+1)|I_{-,A}|^{2},\\
\rho_{eegg}^{41} &  =(2\langle n_{B}\rangle+1)\alpha^{\ast}\beta
I_{-,B}I_{+,B}+\beta^{\ast}\alpha(2\langle n_{A}\rangle+1)I_{+,A}I_{-,A},\\
\rho_{gege}^{22} &  =\alpha^{2}+\beta^{2}|I_{+,A}|^{2}\langle n_{A}%
\rangle(\langle n_{B}\rangle+1)|I_{+,B}|^{2}+\beta^{2}|I_{-,A}|^{2}\langle
n_{A}\rangle\langle n_{B}\rangle|I_{+,B}|^{2}\\
&  \quad+\beta^{2}|I_{-,A}|^{2}(\langle n_{A}\rangle+1)(\langle n_{B}%
\rangle+1)|I_{+,B}|^{2}+\beta^{2}|I_{-,A}|^{2}\langle n_{B}\rangle(\langle
n_{A}\rangle+1)|I_{-,B}|^{2},\\
\rho_{egge}^{32} &  =\alpha^{\ast}\beta+\beta^{\ast}\alpha\langle n_{A}%
\rangle(\langle n_{B}\rangle+1)I_{+,A}I_{+,B}^{\ast}I_{-,A}I_{-,B}^{\ast
}+\beta^{\ast}\alpha\langle n_{A}\rangle\langle n_{B}\rangle I_{+,A}%
I_{-,B}^{\ast}I_{-,A}I_{+,B}^{\ast}\\
&  \quad+\beta^{\ast}\alpha(\langle n_{A}\rangle+1)(\langle n_{B}%
\rangle+1)I_{-,A}I_{+,B}^{\ast}I_{+,A}I_{-,B}^{\ast}+\beta^{\ast}%
\alpha(\langle n_{A}\rangle+1)\langle n_{B}\rangle I_{-,A}I_{-,B}^{\ast
}I_{+,A}I_{+,B}^{\ast},\\
\rho_{geeg}^{23} &  =\alpha\beta^{\ast}+\beta\alpha^{\ast}\langle n_{A}%
\rangle(\langle n_{B}\rangle+1)I_{+,A}^{\ast}I_{+,B}I_{-,A}^{\ast}%
I_{-,B}+\beta\alpha^{\ast}\langle n_{A}\rangle\langle n_{B}\rangle
I_{+,A}^{\ast}I_{-,B}I_{-,A}^{\ast}I_{+,B}\\
&  \quad+\beta\alpha^{\ast}(\langle n_{A}\rangle+1)(\langle n_{B}%
\rangle+1)I_{-,A}^{\ast}I_{+,B}I_{+,A}^{\ast}I_{-,B}+\beta\alpha^{\ast
}(\langle n_{A}\rangle+1)\langle n_{B}\rangle I_{-,A}^{\ast}I_{-,B}%
I_{+,A}^{\ast}I_{+,B},\\
\rho_{egeg}^{33} &  =\beta^{2}+\alpha^{2}|I_{-,A}|^{2}\langle n_{A}%
\rangle(\langle n_{B}\rangle+1)|I_{-,B}|^{2}+\alpha^{2}|I_{-,A}|^{2}\langle
n_{A}\rangle\langle n_{B}\rangle|I_{+,B}|^{2}\\
&  \quad+\alpha^{2}|I_{+,A}|^{2}(\langle n_{A}\rangle+1)(\langle n_{B}%
\rangle+1)|I_{-,B}|^{2}+\alpha^{2}|I_{+,A}|^{2}(\langle n_{A}\rangle+1)\langle
n_{B}\rangle|I_{+,B}|^{2},\\
\rho_{ggee}^{14} &  =(2\langle n_{B}\rangle+1)\alpha\beta^{\ast}I_{-,B}^{\ast
}I_{+,B}^{\ast}+\beta\alpha^{\ast}(2\langle n_{A}\rangle+1)I_{+,A}^{\ast
}I_{-,A}^{\ast},\\
\rho_{eeee}^{44} &  =\beta^{2}(\langle n_{B}\rangle|I_{-,B}|^{2}+(\langle
n_{B}\rangle+1)|I_{+,B}|^{2})+\alpha^{2}(\langle n_{A}\rangle|I_{-,A}%
|^{2}+(\langle n_{A}\rangle+1)|I_{+,A}|^{2}).
\end{align*}
\end{widetext}
where $\langle n\rangle$ represents the expectation value of the particle
number operator in the field state. Due to the characteristic
\textquotedblleft X\textquotedblright\ pattern formed by its non-zero matrix
elements, this state is commonly referred to as an \textquotedblleft
X-state\textquotedblright\ \cite{tanas2004stationary,ali2010quantum}. For
notational simplicity, we have omitted the normalization factor $Tr(\rho
_{out})=\rho_{gggg}^{11}+\rho_{gege}^{22}+\rho_{egeg}^{33}+\rho_{eeee}^{44}$ here.

We use the concurrence \cite{woot1998con} as the method of entanglement
measure for the density operator $\rho_{out}$. The concurrence $C$ of the
matrix $\rho_{out}$ can be calculated using the formula
\cite{tanas2004entangling,yu2005evolution,yu2006sudden}
\begin{equation}
C=2\ast Max\{0,C_{1},C_{2}\},
\end{equation}
where $C_{1}=|\rho_{geeg}^{23}|-\sqrt{\rho_{gggg}^{11}\rho_{eeee}^{44}}$ and
$C_{2}=|\rho_{eegg}^{41}|-\sqrt{\rho_{egeg}^{33}\rho_{gege}^{22}}$.
Substituting the matrix elements from $\rho_{out}$, we have the concurrence of
state (\ref{rho}).

\begin{figure}[ptb]
\includegraphics[width=0.45\textwidth]{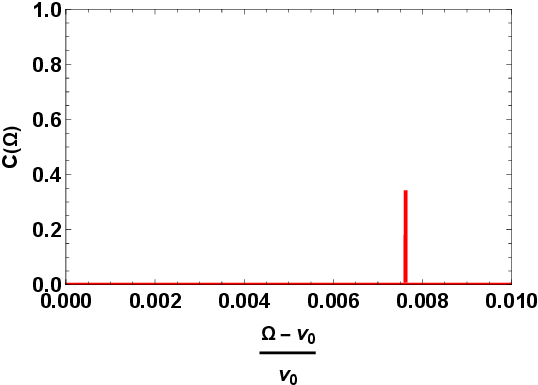} \caption{Entanglement
behavior in thermal fields. The background field temperature $\beta= 0.3$ and
the other parameters are the same as in Fig.1}%
\label{Fig.2}%
\end{figure}

For entangled states undergoing acceleration in thermal fields, the
degradation of entanglement arises from both the thermal radiation of the
background field and the Unruh effect. Figure 2 shows the residual
entanglement after the state has been accelerated in the thermal field. It is
evident that a portion of the entanglement is preserved at a specific energy
level where the AIT effect occurs. This suggests that the AIT effect offers a
degree of protection to entanglement, as the entanglement is almost completely
destroyed at other energy levels due to enhanced detector transitions.

\begin{figure}[ptb]
\includegraphics[width=0.45\textwidth]{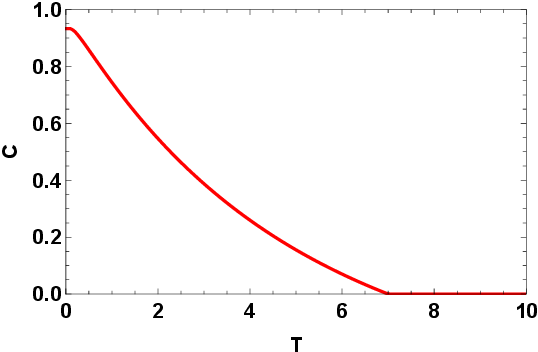} \caption{Residual
entanglement as a function of background field temperature. The detector's
energy gap is fixed at $\Omega=0.00762$, and the other parameters are the same
as in Fig.1}%
\label{Fig. 3}%
\end{figure}

However, this kind of entanglement protection cannot be sustained at all
temperatures. When the temperature of the background thermal field becomes
sufficiently high, the entanglement is still destroyed, even if the conditions
required for the AIT effect are satisfied. This is illustrated in Fig. 3,
which clearly shows how the residual entanglement varies with the field
temperature. As the temperature increases, the residual entanglement decreases
monotonically and nearly vanishes beyond a critical temperature threshold.

\section{Conclusion}

In this work, we investigate the entanglement dynamics of two initially
entangled UDW detectors moving along specially chosen worldlines and analyze
the respective contributions of the Unruh effect and thermal background
fields. These special worldlines are associated with the AIT effect, whereby,
for detectors with specific energy gaps, the stimulated absorption term
becomes negligibly small and the induced excitation by the Unruh effect
dominates, when the background field is non-vacuum. In particular, we show
that the AIT effect persists even in thermal background fields. When two
entangled detectors move under AIT conditions, our results reveal that
although the existence of field particles enhance the detectors' thermal
response and leads to the complete loss of entanglement in general, quantum
entanglement can still survive partially for detectors with specific energy
levels required by the AIT effect. Furthermore, we examine the impact of the
field temperature on entanglement and find that the residual entanglement
enabled by the AIT effect decreases with increasing temperature and vanishes
beyond a critical temperature threshold.

\section{Acknowledgments}

This work is supported by National Natural Science Foundation of China (Grant
Nos. 12375057 and 12047502), and the Fundamental Research Funds for the
Central Universities, China University of Geosciences (Wuhan).

\section{Data availability}

Data sharing is not applicable to this article as no datasets were generated
or analyzed during the current study.

\end{document}